\documentclass{article}
\usepackage{fleqn}
\usepackage{epsfig}
\usepackage{amsmath,amssymb}
\begin{document}

\begin{center}
{\bf\Large Numerical Models of Cosmological Evolution of the Degenerated Fermi-system of Scalar Charged Particles}\\[12pt]
Yu.G. Ignatyev and A.A. Agathonov\\
Kazan Federal University,\\ Kremlyovskaya str., 35,
Kazan 420008, Russia
\end{center}

{\bf keywords}: Early Universe, fantom scalar interaction, relativistic kinetics,
cosmological acceleration, numerical simulation.

\begin{abstract}
Based on mathematical model of the statistical Fermi system with the interparticle interaction which was constructed in the previous articles, this work offers the construction and analysis of the numerical models of cosmological evolution of the single-component degenerated Fermi system of the scalar particles. The applied mathematics package Mathematica 9 is used for the numerical model construction.

\end{abstract}

\newcommand{\Fig}[4]{%
\begin{center}
\parbox{#2cm}{%
\refstepcounter{figure}\includegraphics[width=#2cm,
height=#3cm]{#1} \noindent {\bf Fig. \thefigure.}\quad
#4}\end{center}}
\newcommand{\RFig}[5]{%
\begin{center}
\parbox{12cm}{%
\refstepcounter{figure}\includegraphics[width=#3cm,
height=#2cm,angle=-90]{#1}\vskip #4pt \noindent {\bf Fig.
\thefigure.}\quad #5}\end{center}}
\newcommand{\FRFig}[4]{%
\begin{center}
\parbox{#2.5cm}{%
\refstepcounter{figure}\fbox{\parbox{#2.5cm}{\hskip 2.5mm\vskip 2.5mm
\includegraphics[width=#3cm,
height=#2cm,angle=-90]{#1}}} \vskip 12pt \noindent {\bf Fig.
\thefigure.}\quad #4}\end{center}}
\newcommand{\FFig}[4]{%
\begin{center}
\parbox{#2cm}{%
\refstepcounter{figure}\fbox{\parbox{#2.5cm}{\hskip 2.5mm\vskip 2.5mm
\includegraphics[width=#3cm,
height=#2cm,angle=-90]{#1}}}\vskip 12pt \noindent {\bf –ис.
\thefigure.}\quad #4}\end{center}}

\section{The Mathematical Model}
\subsection{The Mathematical Model Of The Degenerated Scalar Charged Cosmological Plasma}
The mathematical model of the scalar charged particles' statistical
system based on the microscopic description and following procedure of transition to the kinetic theory was formulated in the previous articles  \cite{Ignat14_1,Ignat14_2}.
The strict macroscopic consequences of the kinetic theory are the transport equations including the conservation law of a certain
vector current corresponding to the microscopic conservation law in reactions of a certain fundamental charge ${\rm e}$
(if such a conservation law exists) --
\begin{equation}\label{III.1}
\nabla_i\sum\limits_a {\rm e}_a n^i_a=0,
\end{equation}
as well as the statistical system energy-momentum conservation laws:
\begin{equation}\label{III.2}
\nabla_i T^{ik}_{pl}\equiv \nabla_i \sum\limits_a T^{ik}_{a}=\sigma \nabla ^{i} \Phi,
\end{equation}
where $n^i_a$ is a number vector and $T^{ik}_a$ is an energy-momentum tensor (EMT) of ``a'' - %
sort particles; $\sigma$ is a scalar charge density (see definitions in \cite{Ignat14_2}). %
At conditions of a local thermodynamic equilibrium (LTE) the statistical system is an isotropic one and its
macroscopic moments take form of the corresponding moments of an ideal flux \cite{Ignatev4}:
\begin{equation}\label{III.3}
n^i_a=n_a u^i,
\end{equation}
\begin{equation}\label{III.4}
T^{ik}_a=(\mathcal{E}_{a}+\mathcal{P}_{a}) u^iu^k-\mathcal{P}_{a}g^{ik},
\end{equation}
where $u^i$ is a unit timelike vector of the statistical system's dynamic velocity
\begin{equation}\label{III.5}
(u,u)=1.
\end{equation}
Let us ascertain what are the consequences of the conservation laws (\ref{III.1}) -- (\ref{III.2}) at conditions of the LTE.
From the normalization ratio (\ref{III.5}) it follows the well-known identity law:
\begin{equation}\label{III.6}
u^k_{~,i}u_k\equiv 0.
\end{equation}
With an account of(\ref{III.4}) -- (\ref{III.6}) the conservation laws (\ref{III.2}) can be reduced to form:
\begin{eqnarray}\label{III.7}
(\mathcal{E}_{pl}+\mathcal{P}_{pl})u^i_{~,k}u^k=(g^{ik}\!\!\!-u^iu^k)(\mathcal{P}_{pl,k}+\sigma\Phi_{~,k});\\
\label{III.7a}
\nabla_k(\mathcal{E}_{pl}+\mathcal{P}_{pl})u^k=(\mathcal{P}_{pl,k}+\sigma\Phi_{~,k})u^k,
\end{eqnarray}
and the conservation law of the fundamental charge ${\rm e}$ \ref{III.1}) becomes:
\begin{equation}\label{III.7b}
\nabla_k n_eu^k=0,\quad n_e\equiv \sum\limits_a {\rm e}_a n_a.
\end{equation}
Thus, formally for 3 macroscopic scalar functions $\mathcal{E}, \mathcal{P}, n_e$ and 3 %
independent components of velocity vector $u^i$ the macroscopic conservation laws give us 5 independent
equations (\ref{III.7}) -- (\ref{III.7b})\footnote{equation (\ref{III.7}) is dependant on the other equations
due to the identity (\ref{III.6})}. However not all specified macroscopic scalars
are functionally independent since they all are determined by local equilibrium distribution functions:
\begin{equation}\label{III.8}
f^0_a={\displaystyle \frac{1}{{\rm e}^{(-\mu_a+(u,p))/\theta}}\pm 1},
\end{equation}

where $\mu_a$ is a chemical potential, $\theta$ is a local temperature. If it is resolved the series of the chemical equilibrium conditions when the only one chemical potential remains independent and resolved the equation of the mass surface as well as the scalar potential is given together with the scale factor, the four macroscopic scalars $\mathcal{E}, \mathcal{P}, n_e \sigma$ are determined by two scalars, one of which is a certain chemical potential  $\mu$ and the other is a local temperature $\theta$. This the set of equations (\ref{III.7}) -- (\ref{III.7b}) appears to be the fully determined one.

In the cosmological situation in Friedmann metrics
$$ds^2=dt^2-a^2(t)(dx^2+dy^2+dz^2),$$
all thermodynamic functions depend only on time. It is easy to check that $u^i=\delta^i_4$
turns the equations (\ref{III.7}) into the identities and the set of equations (\ref{III.7a}) -- (\ref{III.7b}) %
is reduced to these two equations:
\begin{equation}\label{III.7a1}
\dot{\mathcal{E}}_{pl}+3\frac{\dot{a}}{a}(\mathcal{E}_{pl}+\mathcal{P}_{pl})=\sigma\dot{\Phi};
\end{equation}
\begin{equation}\label{III.7b1}
\dot{n}_e +3\frac{\dot{a}}{a}n_e=0.
\end{equation}
Thus there remains 2 differential equations in two thermodynamic functions $\mu$ and $\theta$.
At  $\mu\to0$ or $\theta\to0$ limit processing we obtain a set of equations on one function and a problem of contradictoriness of these equations
emerges at that this problem does not depend on presence of a scalar field. We shall show below that this problem is just apparent and
really there are no any contradictions in the set of equations (\ref{III.7a1}) -- (\ref{III.7b1}) emerging even in a case of the degenerated Fermi system.

 The interest for the investigation of the dege\-ne\-rated Fermi system of scalar charged particles is due to the maximum simplicity of the mathematical model and the possibility of interpretation of such a system as a cosmological dark (cold) matter. We will not apply at that any constraints on the value of the particles charge considering including also the situations when this value can be greater than one.

\subsection{The Macroscopic Scalars For The Degenerated single-component Fermi system}
At conditions of full degeneration:
\begin{equation}\label{1}
\theta\to 0.
\end{equation}
the local equilibrium fermion distribution function takes the form of the
step function \cite{Ignatev4}:
\begin{equation}\label{2}
f^0(x,P)=\chi_+(\mu-\sqrt{m_*^2+p^2}),
\end{equation}
where $\chi_+(z)$ is a step Heaviside function.

In this case the result of the macroscopic densities \cite{Ignat14_2} integration is expressed in the elementary functions \cite{Ignatev4}:
\begin{equation}\label{3}
n=\frac{1}{\pi^2}p_F^3;
\end{equation}
\begin{equation}\label{3a}{\displaystyle
\begin{array}{l}
{\rm {\mathcal E}}_{pl} = {\displaystyle\frac{m_*^4}{8\pi^2}}
\bigl[\psi\sqrt{1+\psi^2}(1+2\psi^2)\\[8pt]
-\ln(\psi+\sqrt{1+\psi^2})\bigr];
\end{array}}
\end{equation}
\begin{equation}\label{3b}{\displaystyle
\begin{array}{l}
{\mathcal P}_{pl} ={\displaystyle\frac{m_*^4}{24\pi^2}}
\bigl[\psi\sqrt{1+\psi^2}(2\psi^2-3) \\ [8pt]
+3\ln(\psi+\sqrt{1+\psi^2})\bigr];
\end{array}}
\end{equation}
\begin{equation}\label{3c}{\displaystyle
\begin{array}{l}
\sigma={\displaystyle\frac{q\cdot m_*^3}{2\pi^2}}\left[\psi\sqrt{1+\psi^2}-
\ln(\psi+\sqrt{1+\psi^2})\right];
\end{array}}
\end{equation}
where it is introduced the dimensionless function $\psi$
\begin{equation}\label{psi}\psi=p_F/m_*,
\end{equation}
equal to the relation of the Fermi momentum $p_F$ to the effective mass of fermion.

\subsection{The cosmological model}
Let us consider the formulated earlier self-consistent mathematical model relating to
the cosmological situation for the space-flat Friedman model.
In this case the EMT of the scalar field also takes form of the
ideal isotropic flux EMT:
\begin{equation} \label{MET_s}
T_{s}^{ik} =({\rm {\mathcal E}}_s +{\rm {\mathcal P}}_{s} )v^{i} v^{k} -{\rm {\mathcal P}}_s g^{ik} ,
\end{equation}
where:
\begin{eqnarray}\label{Es}
{\rm {\mathcal E}}_s=\frac{\epsilon_1}{8\pi}(\dot\Phi^2+\varepsilon_2 m_s^2\Phi^2);\\
\label{Ps} {\rm {\mathcal P}}_{s}=\frac{\epsilon_1}{8\pi}(\dot\Phi^2-
\varepsilon_2 m_s^2\Phi^2),
\end{eqnarray}
so that:
\begin{equation}\label{e+p}
{\rm {\mathcal E}}_s+{\rm {\mathcal P}}_{s}=\frac{\epsilon_1}{4\pi}\dot{\Phi}^2.
\end{equation}
The scalar field equation in the Friedman metrics takes the form:
\begin{equation}\label{Eq_S_t}
\ddot{\Phi}+3\frac{\dot{a}}{a}\Phi+\epsilon_2 m^2_s\Phi= -4\pi\epsilon_1\sigma(t).
\end{equation}
The non-trivial Einstein equation is to be appended to these equations:
\begin{equation}\label{Einstein_a}
3\frac{\dot{a}^2}{a^2}=8\pi{\rm {\mathcal E}},
\end{equation}
where ${\rm {\mathcal E}}$ is a summary energy density of the Fermi system and the scalar field. This set of equations describes a closed mathematical evolution model of the degenerated Fermi system with a scalar interaction (см. \cite{Ignat14_2}).

Differentiating the energy density of the Fermi system (\ref{3a}) and taking into account the identity:
\begin{equation}\label{E_P_f}
\mathcal{E}_{pl}+\mathcal{P}_{pl}\equiv \frac{m^4_*}{3\pi^2}\psi^3\sqrt{1+\psi^2},
\end{equation}
let us reduce the energy conservation law for the Fermi system (\ref{III.7a1}) to the form of equation:
\begin{equation}\label{Eq_Pl}
\frac{d}{dt}\ln m_*\psi a=0.
\end{equation}
Hence with an account of the function $\psi$ (\ref{psi}) definition we get:
\begin{equation}\label{ap}
ap_F={\rm Const}.
\end{equation}
This with an account of (\ref{3}) we obtain the fermion number conservation law:
\begin{equation}\label{na3}
a^3n={\rm Const}.
\end{equation}
This notwithstanding the apparent complexity of the equation (\ref{III.7a1}) its solution is easily found: from the Fermi system energy conservation law the particle number conservation law is obtained.

\section{The Numerical Simulation}
\subsection{The Cauchy Problem Definition}
With an account of the integral (\ref{ap}) the problem is reduced to the solution of the set of two differential equations:
1. the first-order equation(\ref{Einstein_a}) -- the Einstein equation and
2. the second-order equation (\ref{Eq_S_t}) -- the field equation. %
To set the Cauchy problem for the system (\ref{Einstein_a})--(\ref{Eq_S_t}) it %
is necessary to set the initial conditions for the values $a(t_{0} )$, $\Phi (t_{0} )$, 
$\dot{\Phi }(t_{0} )$, $p_F(t_{0} )$.  %
Let us hereinafter assume:
\begin{equation} \label{IC} t_{0} =0;\quad a(0)=1;\quad \dot{\Phi }(0)=0. \end{equation}
In the cosmological scenario corresponding to the initial conditions (\ref{IC})
at $t=0$the scalar field kinetic energy is turned to zero and
the scalar field equation of state takes the form:
\begin{equation}\label{EP_s_IC}
\mathcal{P}_s(0) =- 
\mathcal{E}_s(0)= 
- \epsilon_1 \epsilon_2 m^2_s \Phi^2_0.
\end{equation}
Let us note that the first two of the conditions  \eqref{EP_s_IC}  practically determine only the scale units and always can be realized.
  Therefore practically it is necessary to set only the initial conditions for the single sought function  $\Phi (t_{0})$
  and to determine the constant in the relation (\ref{ap}):
\begin{equation}\label{p0}
ap_F=p_0.
\end{equation}
However it is not very convenient for a numerical simulation to set the dimensional functions
 $p_0$ and $\Phi_0$ as the initial conditions. Instead we set two dimensionless scalar functions having the explicit physical meaning:
\begin{eqnarray}
\label{kp0}
\varkappa^0_{pl}= \frac{\mathcal{P}_{pl}(0)}{\mathcal{E}_{pl}(0)}, & \in [0,1/3);\\
\label{etas0}
\eta_S^0=\frac{\mathcal{E}_{S}(0)}{\mathcal{E}_{pl}(0)}, & \in (-\infty,+\infty).
\end{eqnarray}

Setting the relation $\varkappa^0_{pl}$ we can determine the inital Fermi momentum $p_0$ and setting $\eta_S^0$ we can determine $\Phi_0$.  Let us introduce the scalar functions $\varkappa(\mathcal{E})$ needed for the analysis :

\begin{equation} \label{kappa} \mathcal{P}=\varkappa(\mathcal{E})\mathcal{E}\Rightarrow (\mathcal{P}_{s} +\mathcal{P}_{pl} )=\varkappa(\mathcal{E})(\mathcal{E}_{s} +\mathcal{E}_{pl} ) \end{equation}
is a summary barotrope coefficient and  $\Omega (\mathcal{E})$:
\begin{equation} \label{GrindEQ__25_} \Omega =\frac{a\ddot{a}}{\dot{a}^{2} } =-\frac{1}{2} (1+3\varkappa) \end{equation}
is an invariant cosmological acceleration. In such setting the problem is determined by four independent initial conditions -- the nonvarying second and third conditions (\ref{EP_s_IC}), varying (\ref{kp0}) and (\ref{etas0}) ones and also by three parameters -- fundamental constants:  $m,q,m_{s}$. Thus there are 5 arbitrary constants in the problem.

\subsection{The dimensionality of the physical values}

From the effective mass definition as well as the scalar field energy density definition
it follows the dimensionality of these values
in units of the Compton length
\begin{eqnarray}\label{diment}
[t]=l/c\to \ell;  [m]=[\mu]=\hbar/lc\to\ell^{-1}; \nonumber\\
\bigl[\mathcal{E}\bigr] \to \ell^{-4};\bigl[\Phi\bigr] =[m]=[\mu]\to \ell^{-1};[q]\to 1.
\end{eqnarray}
In ordinary units ($[m,l,t]$) the charge $q$ has the dimensionality of $m^{1/2} l^{3/2} t^{-1} $ and the scalar field potential has the one of $[\Phi ]=m^{1/2} l^{1/2} t^{-1} $. Thus in Planck units used in the article the value $q\Phi \sim 1$ corresponds to the effective mass of the scalar charged particles of the Planck mass order.

Further, since at numerical solution of the problem we deal with
the very large numbers, it is necessary to
scale the problem in advance. Let us introduce the dimensionless function instead of the scale factor:
\begin{equation}\label{L0}
\Lambda =\ln a(t);\quad \Lambda(0)=0,                                                                                                              \end{equation}
so that:
\begin{equation} \label{dotL}
\dot{\Lambda }=\frac{\dot{a}}{a} =H(t)
\end{equation}
is a Hubble constant,
\begin{equation}\label{Omega}
\Omega =1+\frac{\ddot{\Lambda }}{\dot{\Lambda }^{2} };
\equiv 1+\frac{\dot{H}}{H^{2}}
\end{equation}
\begin{equation}\label{lps}
\psi=\frac{p_0}{m_*}{\rm e}^{-\Lambda}.
\end{equation}

\subsection{The Normal Set Of Equations}
For the numerical integration of the differential equations set let us bring them to the normal view assuming:
\begin{equation}\label{Z}
Z(t)=\dot{\Phi }
\end{equation}
and resolving the obtained system relative to the derivatives %
$\dot{\Lambda},\dot{\Phi }$ and  $\dot{Z}$, we obtain the normal set of equations:
\begin{eqnarray}
\label{norm1}
\dot{\Lambda}&=& \sqrt{\frac{8\pi}{3}}\sqrt{\mathcal{E}_{pl}+\mathcal{E}_{s}};\\
\label{norm2}\dot{\Phi}&=&Z;\\
\label{norm3}\dot{Z}&=&-3\dot{\Lambda}Z-\epsilon_2m^2_s\Phi-4\pi\epsilon_1\sigma,
\end{eqnarray}
where it is necessary to substitute the expressions for the Fermi system energy density (\ref{3a}) and scalar field (\ref{Es}) in the equation (\ref{norm1}) with an account of the function $\psi$ (\ref{lps}) definition and equation (\ref{norm2}); the equation (\ref{norm3}) should have substituted in it the expression for $\dot{\Lambda}$ from the obtained equation (\ref{norm1}) and the expression for $\sigma$ obtained from the relation (\ref{3c}).

\section{The Numerical Simulation Results}
The numerical integration of the set of equations (\ref{norm1})--(\ref{norm3}) was carried out in the applied mathematics package ``Mathematica 9''. Since the set of differential equations reveals the signs of stiffness, there was used the numerical method with an automatic switch from the method ``stiff' to the precise Runge-Kutta method in form of
\begin{verbatim}Method ->{StiffnessSwitching,
Method ->{ExplicitRungeKutta,Automatic}},
AccuracyGoal -> 20, PrecisionGoal -> 5,
MaxSteps ->20000
\end{verbatim}
with a maximum number of steps equal to 20000. Below there are shown the certain results of the simulation \footnote{not more than 5\% from the total number.}. For the convenience of classification we speak about the scalar field as of the field of bosons besides in case of  $m_s\ll 1$ we speak about light boson and in case of $m_s\backsimeq 1$ we speak about heavy bosons. The same is for fermions: $m\ll 1$ are light fermions, $m\backsimeq 1$ are heavy ones; also  $q\backsimeq 1$ are heavy fermions because of their potential energy. Next, in case of $\eta_S\ll 1$ we speak about fermion dominated system, in case of  $\eta_s\gg 1$ we speak about boson dominated system. Everywhere in the presented plots it is  $m_s=0,001$ and the classification is carried out by the initial values $\varkappa^0_{pl}$, $\eta^0_S$.

\subsection{The Ultrarelativistic Fermions $\varkappa^0_{pl}=1/3; m=0.01$,  %
The Fantom Scalar Field, The Boson Dominated System: $\eta^0_S=100$}
In this case there are obtained the following results:
\Fig{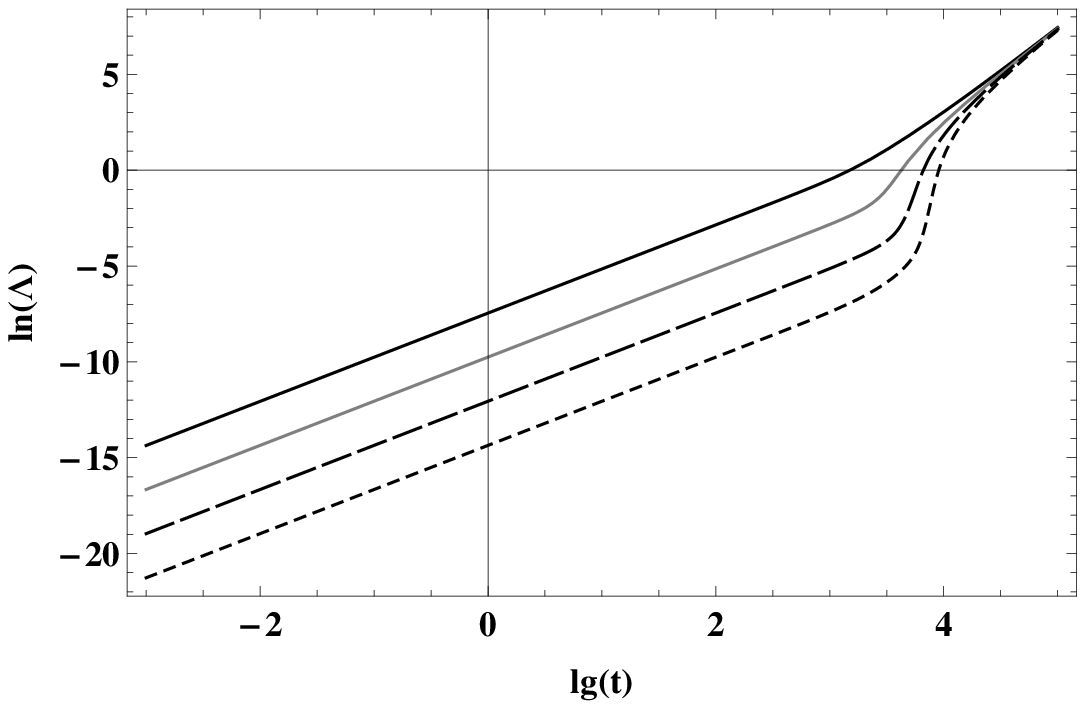}{8.5}{6}{The dependency of the scale function $\lg\Lambda(t)$ evolution on the fermion scalar charge value.
Top-down: $q=0.001;0.1;1;10$. }
The next approximate initial values of the Fermi momentum $p_F^0$ and the scalar potential $\Phi_0$
correspond to the given initial parameters and constants:
\[\begin{array}{ll}
\{q,p^0_F,\Phi_0\}: & \{0.001,0.0112, -0.99\}, \\
\{0.1,0.0035,-0.100\}, & \{1,0.0011, -0.0100\}, \\
\{10,0.00035, -0.00100\}. & \\
\end{array}
\]
\Fig{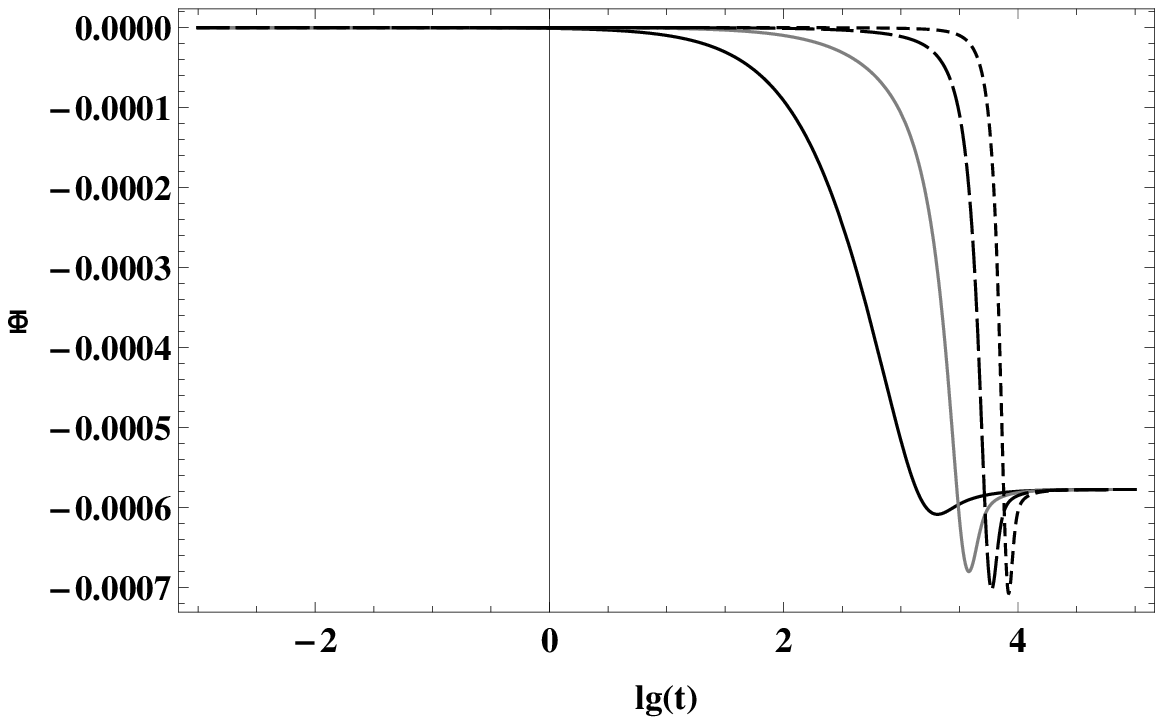}{8.5}{6}{The dependency of the potential derivative $\dot{\Phi}$ evolution on the fermion scalar charge value. Left to right it is: $q=0.001;0.1;1;10$. }

On the plot it is shown the dependency of the invariant cosmological acceleration
$\Omega$ on the scalar charge value. The presence of ``phantom stalagmites'' being a characteristic burst of the acceleration
at times of order $3\div 5\cdot10^3\ t_{pl}$  is itself the feature (Fig. \ref{O1}). Plots $\varkappa(t)$, on the contrary, contain ``fantom stalagmites'' at the same evolution times. It is necessary to notice that given phantom emissions are not the results of the numerical calculations errors. This fact had been checked repeatedly in different models and at different accuracy of calculations. Specified phantom stalagmites correspond to phantom stalactites on the plots of scalar potential.

\Fig{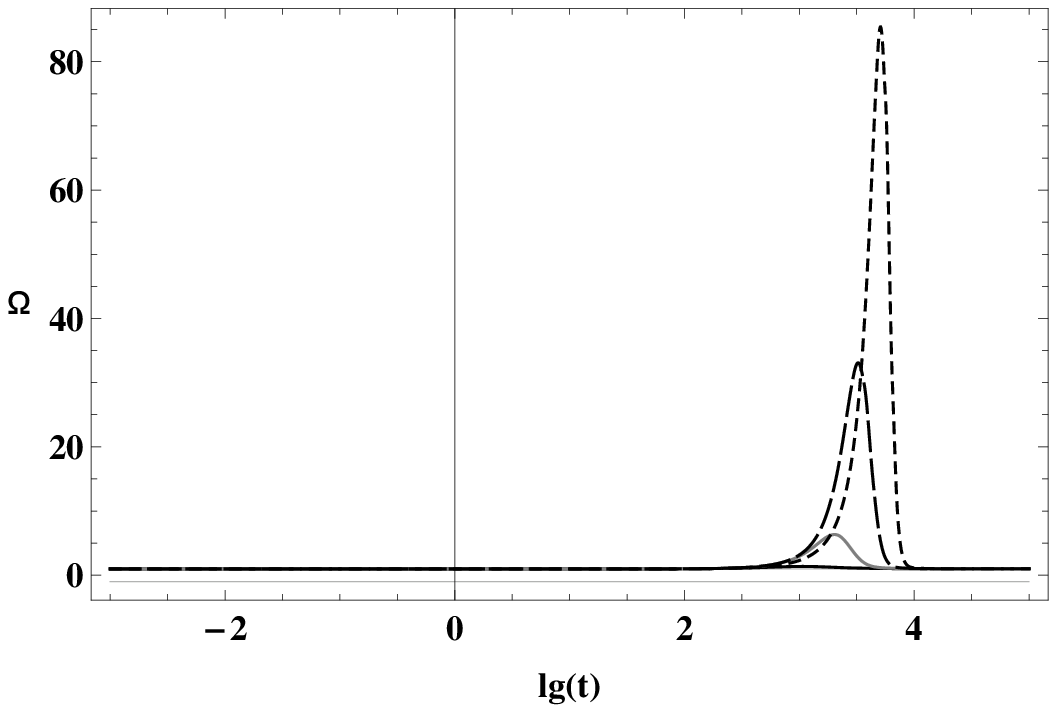}{8.5}{6}{\label{O1}The dependency of the invariant cosmological acceleration %
$\Omega$ evolution on the value of the fermion scalar charge. Bottom-up: $q=0.001;0.1;1;10$. }
On Fig. \ref{lgEsEp1} it is shown the dependency of the summary energy density  $\mathcal{E}$ evolution on %
the value of the fermion scalar charge,
\Fig{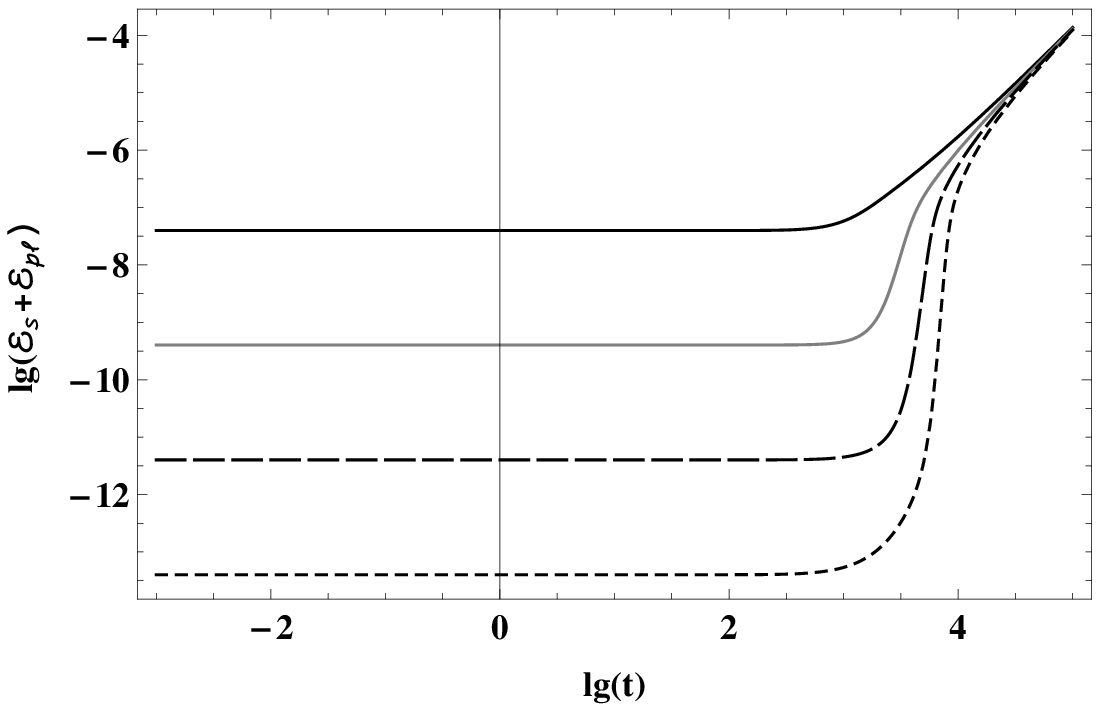}{8.5}{6}{\label{lgEsEp1}The dependency of the invariant summary energy density $\mathcal{E}$
evolution on the value of the fermion scalar charge. Top-down it is: $q=0.001;0.1;1;10$. }
and on the Fig. \ref{etaS1} is shown the dependency of the parameter $\eta_S$ (\ref{etas0}) evolution on the
value of the fermion scalar charge.
\Fig{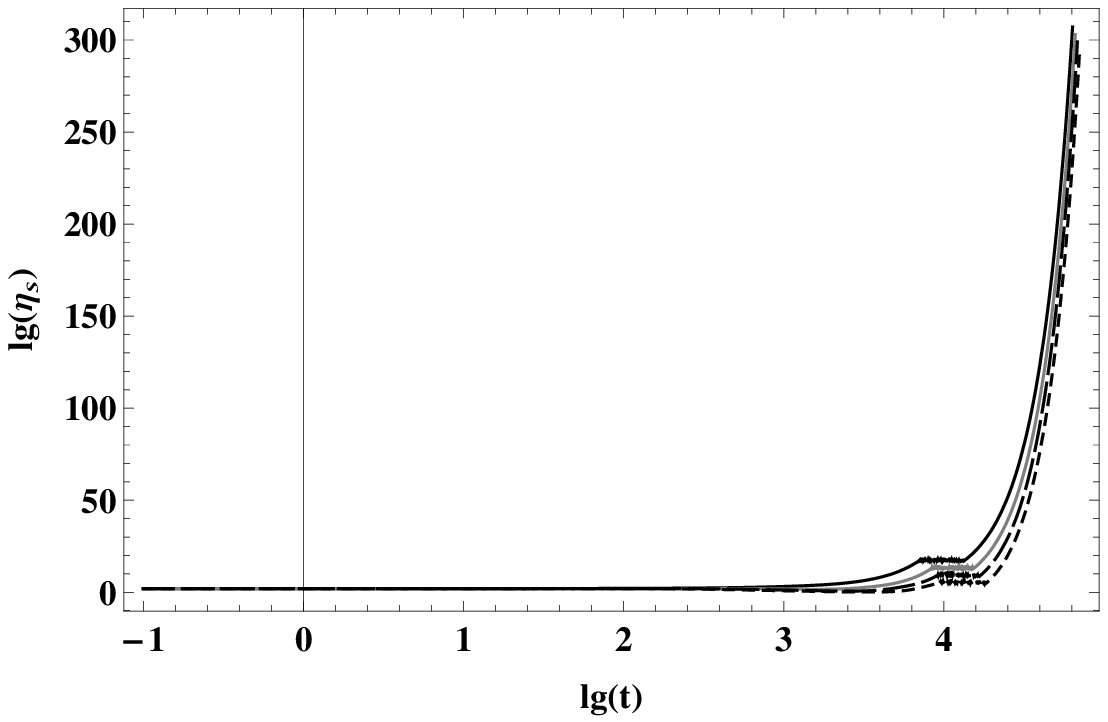}{8.5}{6}{\label{etaS1}The dependency of the parameter $\eta_S$ evolution on the value of the fermion scalar charge.
Top-down it is: $q=0.001;0.1;1;10$.}

From these plots one can see that, first of all,the plot of the summary energy density comes to the common asymptote at variables $(\lg t,\lg\mathcal{E})$, which corresponds to the summary energy density power law. Secondly, one can see that the relation of the scalar field energy density modulus to the Fermi system energy density grows rapidly reaching huge values of $\eta_S\sim 10^{300}$ at $t\sim 10^5$! The given example shows that at large times $t>10^4$ the model practically does not depend on the value of charge and reveals the behavior close to the behavior of the minimal model. However at $t<10^4$ the model behavior significantly depends on the scalar charge value.

\subsection{The Ultrarelativistic Fermions $\varkappa^0_{pl}=1/3; m=0.1$,  %
The Fantom Scalar Field, The Fermion Dominated System: $\eta^0_S=0.01$}
The following approximate initial values of the Fermi momentum $p_F^0$ and the scalar potential $\Phi_0$ correspond to the set-up initial parameters and constants:
$[[q,p^0_F,\Phi_0]$:\\ $[[0.001,1.09, -95.5]$, $[0.1,0.11, -1.00]$,\\ $[1,0.035, -0.100]$, $[10,0.011, -0.010]].$

In this case the following results are obtained (figure \ref{L2}).
On the figure \ref{dotPhi} it is shown the dependency of the invariant cosmological acceleration $\Omega$ evolution on the value of the scalar
charge.

On Fig. \ref{lgEsEp2} it is shown the dependency of the summary energy density $\mathcal{E}$ evolution on the value of the fermion
scalar charge.

\Fig{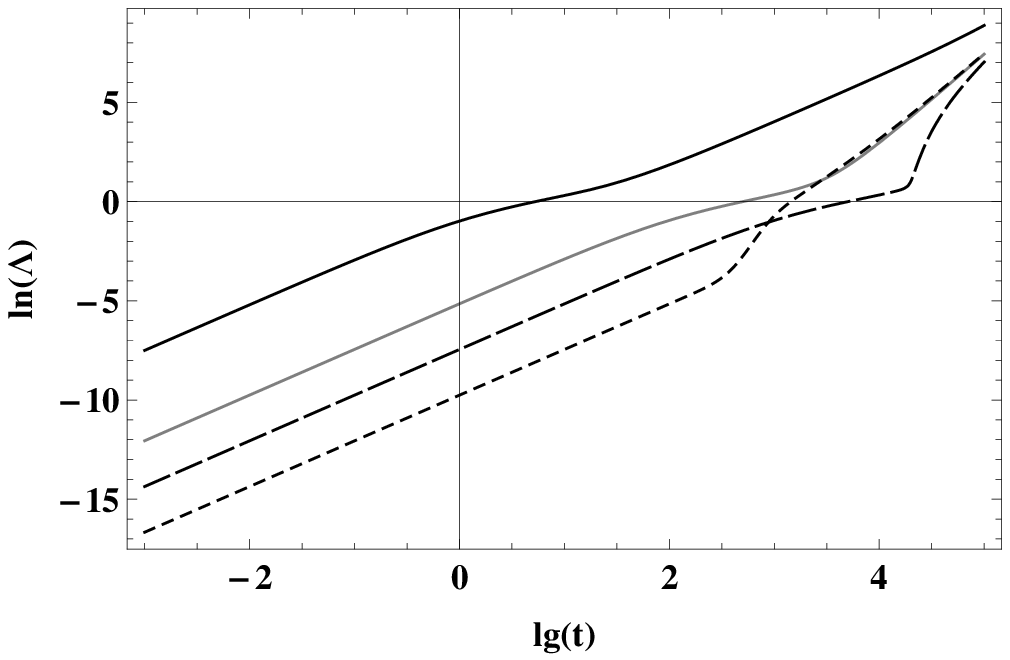}{8.5}{6}{\label{L2}The dependency of the scale function $\lg\Lambda(t)$ evolution%
on the value of the fermion scalar charge. Top-down it is: $q=0.001;0.1;1;10$. }
\Fig{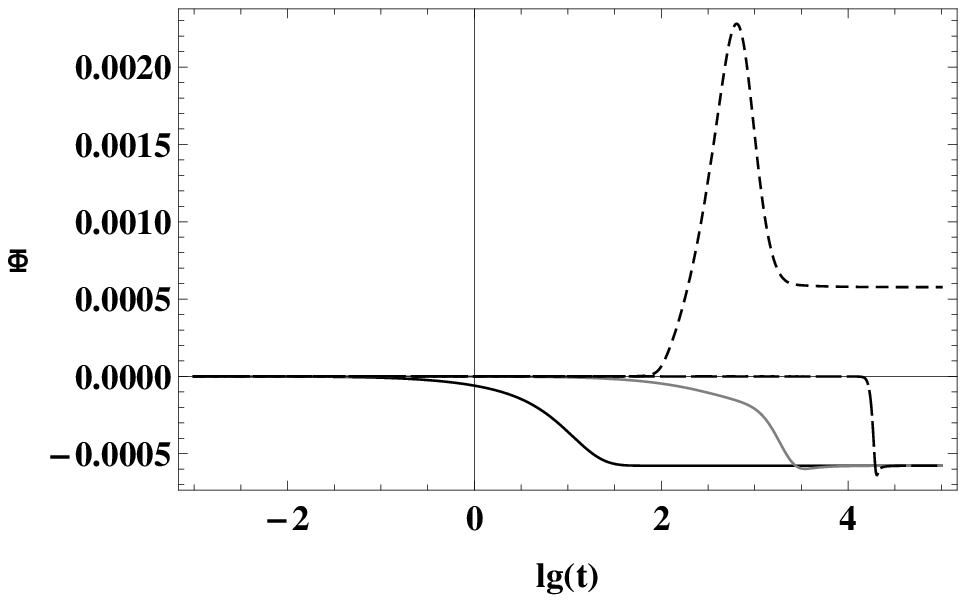}{8.5}{6}{\label{dotPhi} The dependency of the potential derivative $\dot{\Phi}$ evolution on the value of the fermion scalar charge. Left to right it is: $q=0.001;0.1;1;10$. }

\Fig{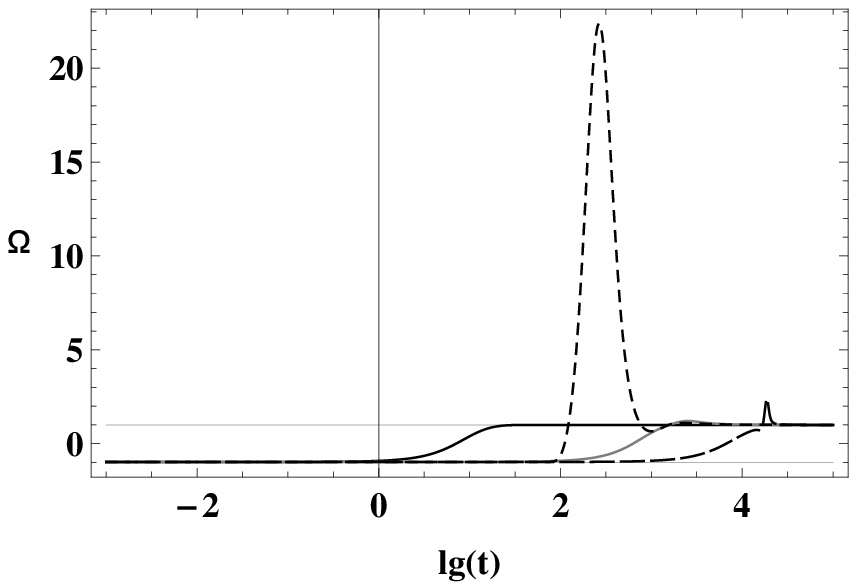}{8.5}{6}{\label{O2}The dependency of the invariant cosmological acceleration $\Omega$ evolution on the value of the scalar
charge. Bottom-up it is: $q=0.001;0.1;1;10$. }
\subsection{The Comparison With A Case Of The Scalar Field. The Relativistic Fermions $p^0_F=3; m=1, m_s=0.1$,  %
$\Phi_0=0.35$}
Let us present the results of the  comparison for the systems with the classic and the phantom scalar fields. In this case we will take the same initial conditions: Fig. \ref{L3} --\ref{etaS3}.

\Fig{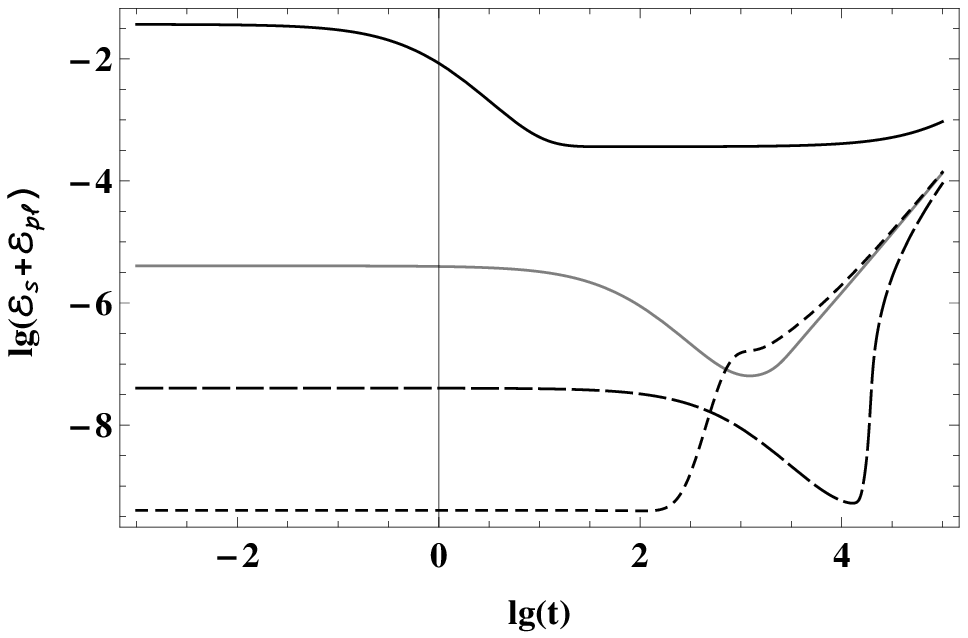}{8.5}{6}{\label{lgEsEp2}The dependency of the invariant summary energy density $\mathcal{E}$
evolution on the value of the fermion scalar charge. Top down it is: $q=0.001;0.1;1;10$. }
and on the Fig. \ref{etaS2} is shown the dependency of the parameter $\eta_S$ (\ref{etas0}) evolution on the
value of the fermion scalar charge.
\Fig{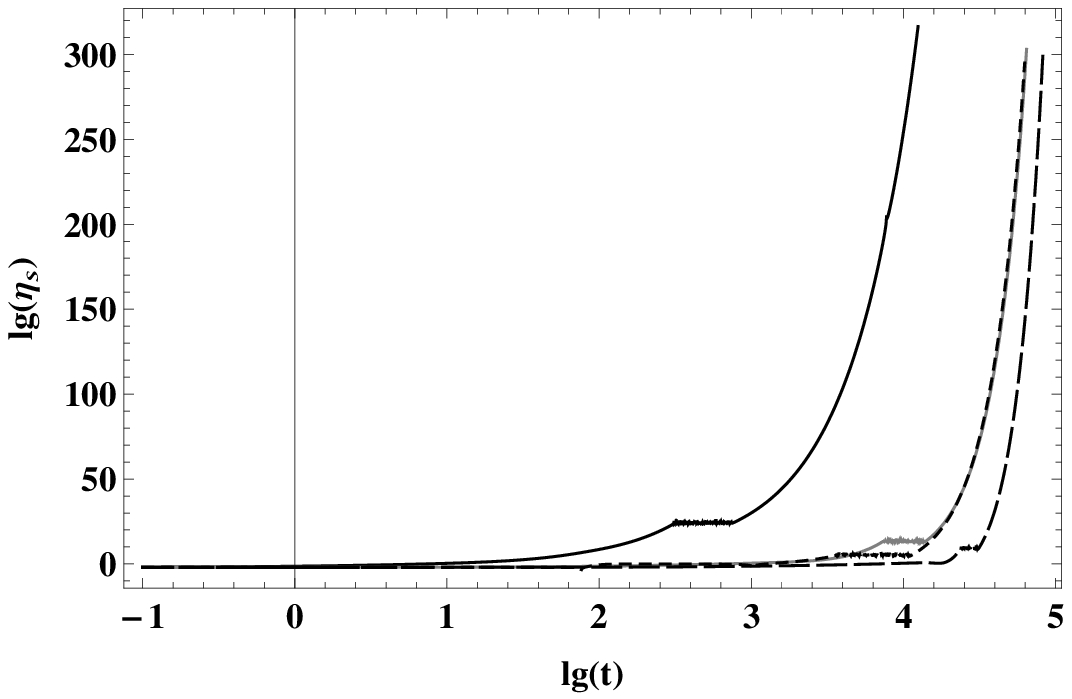}{8.5}{6}{\label{etaS2}The dependency of the parameter $\eta_S$ %
evolution on the
value of the fermion scalar charge. Top down it is: $q=0.001;0.1;1;10$.}

\Fig{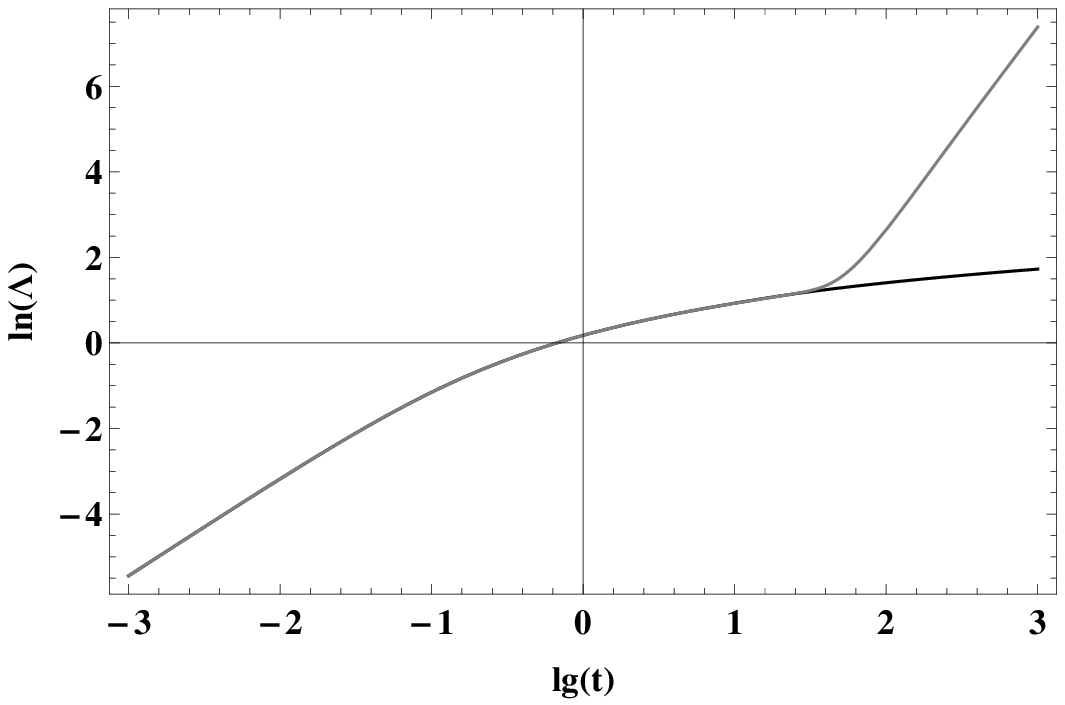}{8.5}{6}{\label{L3}The dependency of the scale function $\lg\Lambda(t)$ evolution on the
value of the fermion scalar charge. The grey line shows the phantom field and the black line shows the classic field. }
\section*{The Conclusion}
Thus we can state that at large times $\sim 10^5 t_{Pl}$ the cosmological evolution of matter based on the Fermi system of scalar charged particles with a phantom interaction does not vary from the evolution of matter with a minimal scalar interaction. However at smaller times $10^5 t_{Pl}$ the evolution of matter with a nonminimal scalar interaction is characterized by the sufficiently greater diversity of the behavior types as compared to the matter with a minimal scalar interaction and also by the presence of the phantom bursts. As such, in contrast to the system with a classical scalar interaction in the system of fermions with a phantom scalar interaction the microscopic oscillations that could lead to the system heating and born of the secondary particles are not emerged. Let us also notice that here is described quite a small part of the obtained results of the numerical simulation. The separate article will be devoted to the systematic description and analysis of these results.

\Fig{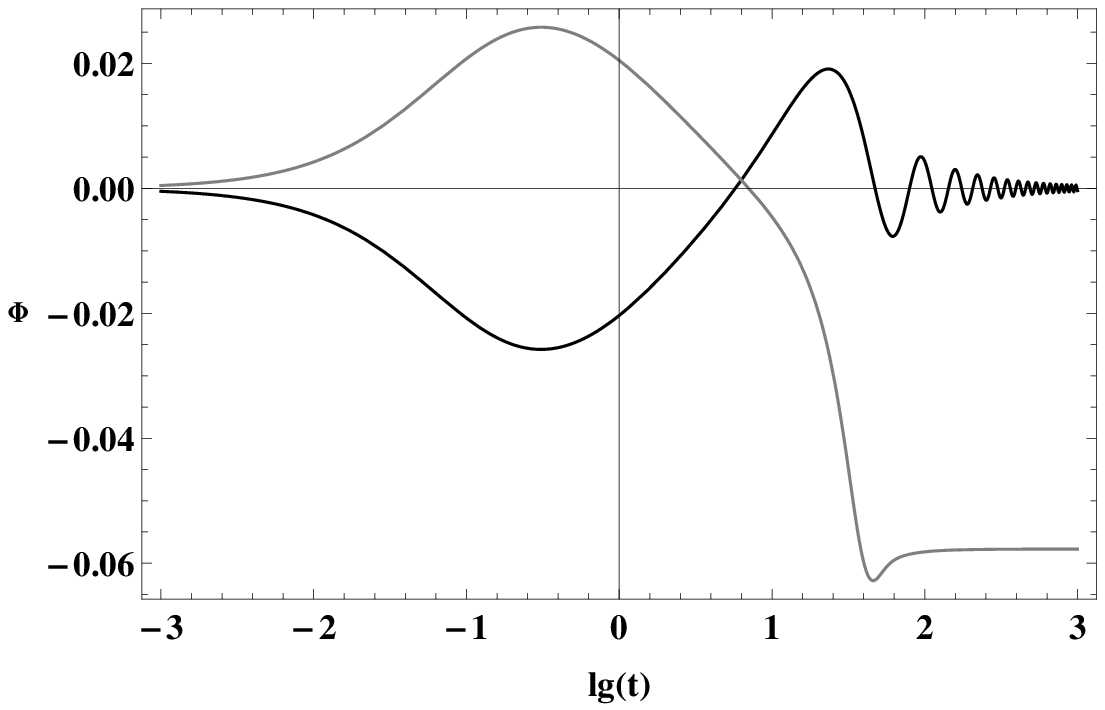}{8.5}{6}{\label{dF3}The dependency of the potential derivative $\dot{\Phi}$ evolution on the value of the fermion scalar charge. The grey line denotes the phantom field and black line denotes the classic field. }

\Fig{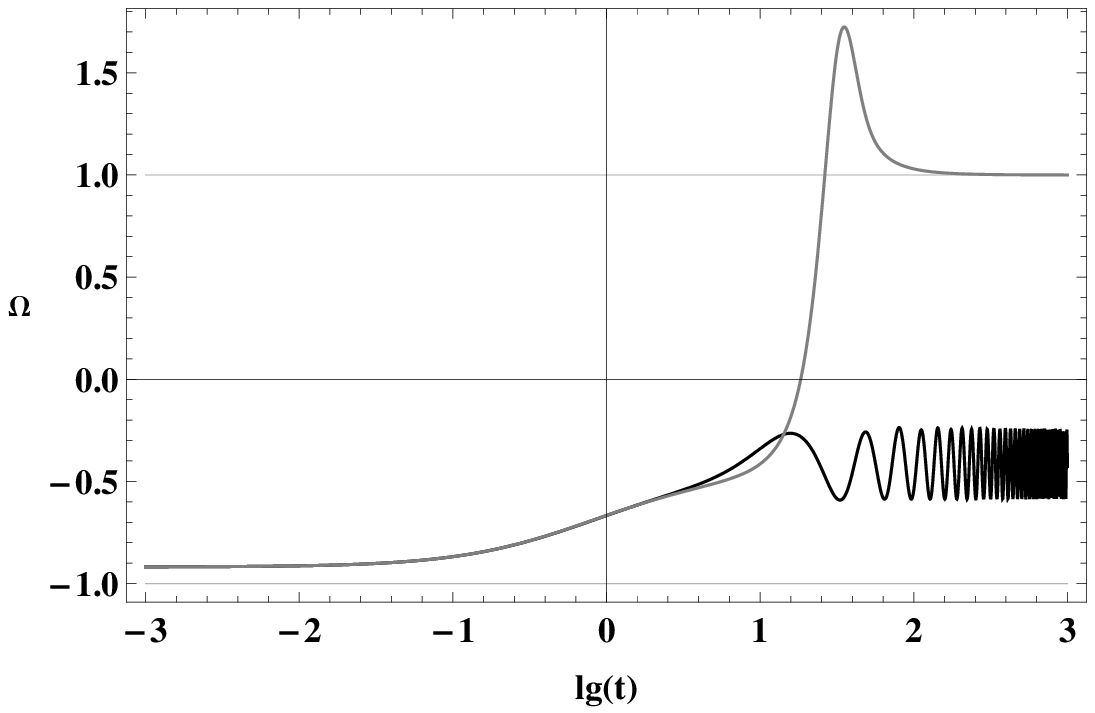}{8.5}{6}{\label{O3}The dependency of the invariant cosmological acceleration $\Omega$ evolution on the value of the scalar
charge. The grey line denotes the phantom field and black line denotes the classic field. }
On Fig. \ref{etaS3} it is shown the dependency of the parameter $\eta_S$ %
evolution on the
value of the fermion scalar charge.
\Fig{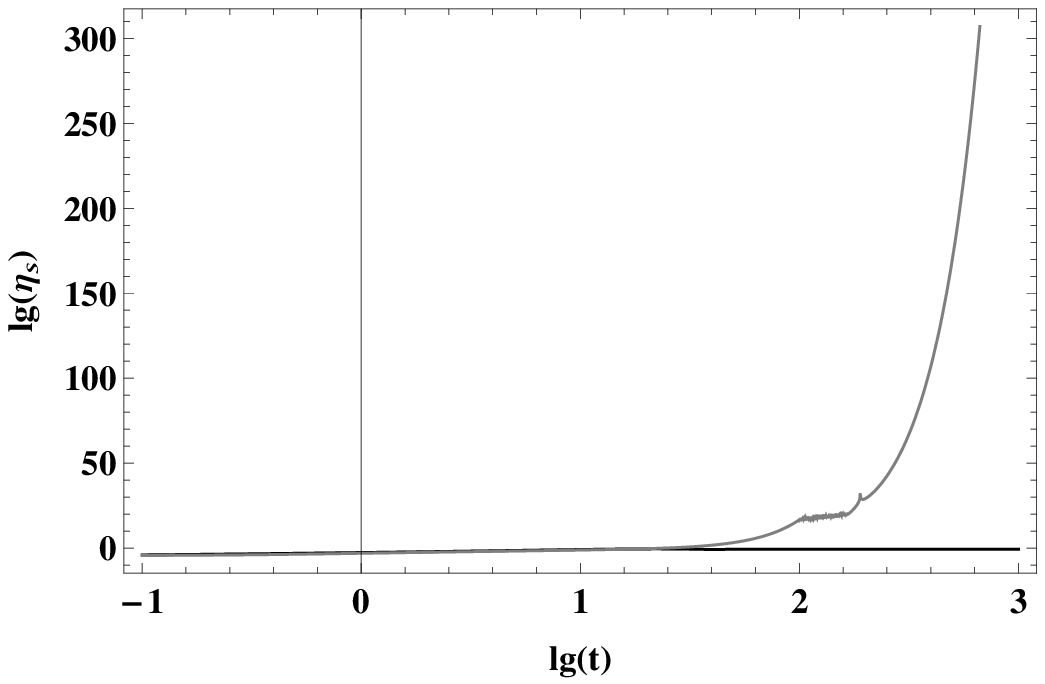}{8.5}{6}{\label{etaS3}The dependency of the parameter $\eta_S$ %
evolution on the
value of the fermion scalar charge. The grey line denotes the phantom field and black line denotes the classic field.}

\end{document}